# Machine Learning-Enabled IoT Security: Open Issues and Challenges Under Advanced Persistent Threats

ZHIYAN CHEN, University of Ottawa, Canada
JINXIN LIU, University of Ottawa, Canada
YU SHEN, University of Ottawa, Canada
MURAT SIMSEK, University of Ottawa, Canada
BURAK KANTARCI, University of Ottawa, Canada
HUSSEIN T. MOUFTAH, University of Ottawa, Canada
PETAR DJUKIC, Ciena, Canada

Despite its technological benefits, Internet of Things (IoT) has cyber weaknesses due to the vulnerabilities in the wireless medium. Machine learning (ML)-based methods are widely used against cyber threats in IoT networks with promising performance. Advanced persistent threat (APT) is prominent for cybercriminals to compromise networks, and it is crucial to long-term and harmful characteristics. However, it is difficult to apply ML-based approaches to identify APT attacks to obtain a promising detection performance due to an extremely small percentage among normal traffic. There are limited surveys to fully investigate APT attacks in IoT networks due to the lack of public datasets with all types of APT attacks. It is worth to bridge the state-of-the-art in network attack detection with APT attack detection in a comprehensive review article. This survey article reviews the security challenges in IoT networks and presents the well-known attacks, APT attacks, and threat models in IoT systems. Meanwhile, signature-based, anomaly-based, and hybrid IDSs are summarized for IoT networks. The article highlights statistical insights regarding frequently applied ML-based methods against network intrusion. Finally, open issues and challenges for common network intrusion and APT attacks are presented for future research.

CCS Concepts: • **Security and privacy** → **Intrusion detection systems**; **Network security**; • **Computing methodologies** → **Machine learning**; • **Networks** → **Sensor networks**.

Additional Key Words and Phrases: Internet of Things, cyber-attack, network security, network intrusion detection, machine learning, deep learning

Authors' addresses: Zhiyan Chen, zchen241@uottawa.ca, University of Ottawa, Ottawa, ON, K1N 6N5, Canada; Jinxin Liu, jliu@uottawa.ca, University of Ottawa, Ottawa, ON, K1N 6N5, Canada; Yu Shen, yshen041@uottawa.ca, University of Ottawa, Ottawa, ON, K1N 6N5, Canada; Murat Simsek, murat.simsek@uottawa.ca, University of Ottawa, Ottawa, ON, K1N 6N5, Canada; Burak Kantarci, burak.kantarci@uottawa.ca, University of Ottawa, Ottawa, ON, K1N 6N5, Canada; Hussein T. Mouftah, mouftah@uOttawa.ca, University of Ottawa, Ottawa, ON, K1N 6N5, Canada; Petar Djukic, pdjukic@ciena.com, Ciena, Ottawa, ON, K2K 0L1, Canada.

## 1 INTRODUCTION

The Internet of Things (IoT) is a network of devices equipped with sensing, communicating, and unique identifiable characteristics [73]. The widespread adoption of connected services that utilize IoT devices and networks has enabled IoT applications' advancement. IoT networks offer smarter environments and services than earlier networks by leveraging and integrating sensed data from various sensors [73]. Embedded sensors in IoT devices communicate with peer nodes through wireless networks such as the standards developed by the 3rd Generation Partnership Project (3GPP), IEEE 802.11ah, Bluetooth Low Energy, and Z-Wave protocols [9]. Additional benefits include transparency and ease of access to information while reducing human interventions.

Due to a number of advantages of IoT, such as improved customer engagement, reduction of waste, enhanced data collection, and technology optimization, IoT technology is being integrated into the various system (e.g., industrial control systems, modern vehicles, and Critical Infrastructure (CI)) [133]. For instance, the integration of IoT systems with CIs such as facilities, power distribution networks, and critical assets is emergent [23]. One such example is the use of IoT systems to monitor and manage CIs in the smart city context [157]. Such networks' objective is to improve efficiency in management, production, and services of a city to make the city more livable.

Another set of IoT use cases is in industrial settings. Industrial IoT (IIoT) is a subset of IoT, which is the application of IoT applied to industrial processes by facilitating the interconnection of entities such as controllers, sensors, and actuators in industrial production and automation context [177]. IIoT is utilized in CIs that provide many essential services in the industrial setting[103]. IIoT contains closely related IoT concepts, including the same main characteristics: availability, intelligent control, and connected devices. Therefore, IoT and IIoT share some common concepts, including data connectivity, data security, secure cloud. However, IIoT is generally used to monitor and control industrial processes such as continuous manufacturing, supply chain monitoring, and management systems. IIoT also uses more sensitive and precise sensors, including more location-aware technologies than IoT [148].

This survey focuses on network attacks in both broad applications of IoT systems. Both general IoT and IIoT systems are vulnerable to a variety of cyberattacks. In the case of industrial control systems, safety is as critical as the notion of security. Safety addresses such concerns as ensuring the protection of humans, the environment, and equipment against consequences of system failures [186]. To understand the importance of some critical industrial infrastructure, think of a nuclear plan using IIoT and the consequences of a cyberattack on this infrastructure. An attack could have consequences on the personnel at the plant, and it could have broad-ranging effects on the world.

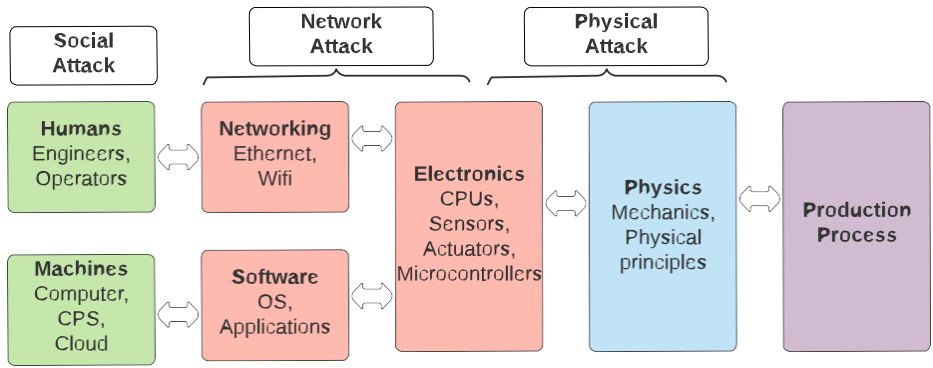

Fig. 1. Attack surface in a CPPS [133]

As an example, we use the architecture of a Cyber-Physical Production System (CPPS) (Fig. 1), which may be a part of an IIoT-based smart factory. The architecture consists of several layers [133]. The system's base is the actual production process - the factory consisting of the physical and production components. The IIoT system is built on top and consists of electronics, networking, and software. Finally, humans and machines monitor the system to ensure it is working correctly.

This IIoT system is vulnerable across multiple layers. For instance, physical layers are susceptible through invasive hardware attacks [174] and side-channel attacks. Meanwhile, social attacks (e.g., phishing attacks) arise when people operate a CPPS.

Since IoT-based systems operate on wireless channels that broadcast data, they are potentially exposed to malicious users, and cybercrimes [186]. Considering the physical medium, software, and protocols, vulnerabilities combined with adversaries' competency can produce sophisticated network-based attacks on IoT systems [93].

There are various motives of hackers in attacking an IoT system, including exposing, altering, disable, destroying, stealing, or gaining unauthorized access to or making unauthorized use of an asset [115]. An example is a Denial- of-Service (DoS) in which an attacker floods messages to victims by sending a large number of packets from a legitimate host network [135]. Alternatively, port scanning attacks [79] involve an attacker that detects an open port where attacks can be injected. Consequently, attackers can obtain network-related detailed information, including but not limited to IP and MAC addresses. By the end of 2021, over 10 billion IoT devices are employed in smart city and industrial management on a global scale. This number will approach 75 billion by the end of 2025 according to the study in [129]. Due to the enormous expansion of the IoT devices in recent years and the vulnerability of IoT sensors, governments and businesses invest on IoT security to prevent hackers from infiltrating, and the investments to keep IoT devices secure is still insufficient to prevent hackers from causing economic losses of more than 1 trillion US dollars [124]. These figures demonstrate the critical significance of IoT security and its pervasive impact on daily life and economic activity.

Detecting and mitigating network intrusion is essential to secure an IoT system. For instance, as reported in [78], IoT-based smart grid networks, with millions of nodes, creates a large attack surface. The protection of IoT systems relates primarily to sustaining the essentials for daily needs of societies and individuals [20]. In hybrid detection systems, ML based identification methods have been researched intensively and have been shown to operate efficiently against network intrusion detection. State of the art today contains various ML methods, some of which exhibit competitive detection levels reaching 100% [99].

An advanced persistent threat (APT) intends to bypass access control mechanisms so to control user equipment. APT attacks are crucial for users due to their long time span until damage, and are challenging for identification in a network due to hidden characteristics [54]. APT attacks are essential parts of network datasets. For instance, NSL-KDD contains U2R and R2L classes, which belong to APT attacks [68]. Moreover, backdoor, reconnaissance, Shellcode, and worms in UNSW-NB15 are also among APT attacks. We summarize renown datasets and classes in Table 1, which shows datasets include various types of APT attacks. Furthermore, there exists no public network dataset with only APT attacks. It is difficult to investigate APT attacks using one dataset; hence the literature that focuses on APT attacks is limited. Typically, APT attacks occupy extremely lower percentage (e.g., 0.046% U2R and 0.88% U2R in NSL-KDD) in order to disguise their patterns. Therefore, APT attacks detection performance is lower than non-APT attacks. With this in mind, this article aims to contribute to the IoT security literature by reviewing the existing research on APT attacks and presenting opportunities, challenges and open issues for further research.

Based on the efficiency of ML-based approaches in the detection of widely investigated attack types in IoT, this paper aims to review these attack models in IoT-monitored CIs and review the use of ML algorithms to detect those attacks. Meanwhile, it is worthy to review APT attacks for future prospective directions.

The contributions in the survey are as follows:

Table 1. Attacks in public datasets. ✓ means attack in dataset; x means does not belong to dataset.

| | Attack | NSL-KDD | KDD99 | UNSW-NB15 | MAWLab | CICIDS2018 | AWID |
|---|---|---|---|---|---|---|---|
| APT Attack Patterns | Probe | ✓ | ✓ | x | x | x | x |
| | R2L | ✓ | ✓ | x | x | x | x |
| | U2R | ✓ | ✓ | x | x | x | x |
| | Backdoors | x | x | ✓ | x | x | x |
| | Exploits | x | x | ✓ | x | x | x |
| | Reconnaissance | x | x | ✓ | x | x | x |
| | Shellcode | x | x | ✓ | x | x | x |
| | Worms | x | x | ✓ | x | x | x |
| | Botnet | x | x | x | x | ✓ | x |
| | Web Attacks | x | x | x | x | ✓ | x |
| | Infiltration | x | x | x | x | ✓ | x |
| | IPV6 Tunneling | x | x | x | ✓ | x | x |
| | NetworkScan ICMP, UDP, TCP | x | x | x | ✓ | x | x |
| | PortScan | x | x | x | ✓ | x | x |
| | Injection | x | x | x | x | x | ✓ |
| | Impersonation | x | x | x | x | x | ✓ |
| Non-APT Attack Patterns | DoS | ✓ | ✓ | x | ✓ | ✓ | |
| | Fuzzers | x | x | ✓ | x | x | x |
| | Analysis | x | x | ✓ | x | x | x |
| | DoS | x | x | ✓ | x | x | x |
| | Generic | x | x | ✓ | x | x | x |
| | Burte-force | x | x | x | x | ✓ | x |
| | Heartbleed | x | x | x | x | ✓ | x |
| | DDos | x | x | x | x | ✓ | x |
| | HTTP | x | x | x | ✓ | x | x |
| | Multi. Points | x | x | x | ✓ | x | x |
| | Alpha | x | x | x | ✓ | x | x |
| | Flooding | x | x | x | x | x | ✓ |

- Presents the architectures of IoT-monitored CIs and their security vulnerabilities.
- Reviews various AI-based state-of-the-art approaches to detect and mitigate network intrusions. Out of these, 14 AI methods are shortlisted based on their frequency of application.
- Public IoT datasets are summarized for attacks detection researching which are barely concluded in current literature.
- Opportunities, open issues and challenges against APT attacks, which exist in most current network detection datasets and typically result in low detection performance.

Recently, several surveys discussed ML approaches for network intrusion as shown in Table 2. Although ML-based approaches have emerged, it is valuable to study the legacy methodologies to understand their benefits and limitations. Moreover, APT attacks exist in many public datasets but they are rarely studied in surveys highlighting the APT attack detection via ML. Meanwhile, an IoT-specific dataset is an essential part for IoT security research using machine learning algorithms. However, it is difficult and expensive to deploy IoT networks to collect data samples being an IoT dataset. Using a public IoT-targeted dataset is a cost effective approach to set up a project and focus on methodologies. However, there is rare literature summarizing IoT-specific information. With these motivations, this article aims to bridge these gaps, 1) including studying traditional techniques and ML-based approaches for attack detection in one paper, 2) reviewing focusing on APT attacks detection via ML approaches, and 3) summarizing existing public IoT datasets. In Table 2, gap analysis with respect to the existing relevant surveys is shown based on the following criteria: ML and/or Deep Learning (DL), legacy IDS techniques, attack and threat models, and APT consideration. Meanwhile, abbreviations in this work are summarized in Table 3.

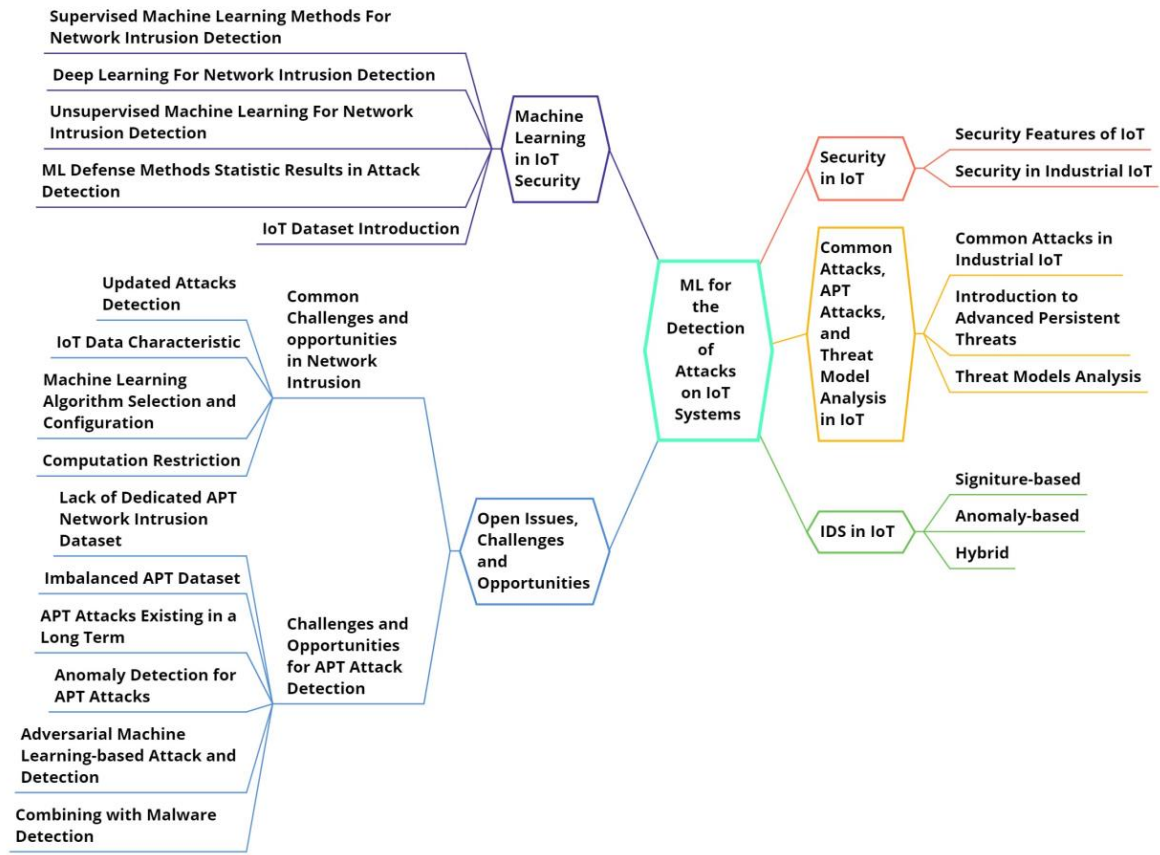

Fig. 2. Survey Structure

The structure of this survey is shown in Fig. 2, and the rest of the survey is organized into the following six sections: Sec. 2 presents security in IoT, including security features and specific security for IIoT (e.g., Industrial Control Systems); Sec. 3 discusses IoT attacks by also focusing on the APT attacks. Moreover, an attack taxonomy is presented with also APT attacks distribution. Several threat models are reviewed and an attack analysis is presented by using the Process for Attack Simulation and Threat Analysis (PASTA) model. Sec. 4, network intrusion detection methodologies for IoT are reviewed and categorized into three groups. Sec. 5 presents the ML-based approaches that are applied to address network attacks; Sec. 6 introduces open issues, challenges, and opportunities for common network intrusion and APT using ML algorithms; Finally, Sec. 7 concludes and summarizes the survey.

Table 2. Comparison with related surveys

| Ref. | ML/DL | Non-M Techniques | Threat Model | APT | Limitations |
|------|-------|------------------|--------------|-----|-------------|
| [51] | DL | ✗ | ✗ | ✗ | DL-based detection methods without ML |
| [76] | Both | ✓ | ✗ | ✗ | No threat analysis specific to the IoT network attack pattern |
| [41] | Both | ✗ | ✗ | ✗ | No discussion of non-ML-based detecting techniques and no traditional security methods |
| [32] | Both | ✓ | ✓ | ✗ | No APT attack life cycle and attack pattern discussion |
| [60] | Both | ✗ | ✗ | ✗ | No network attacks categorized and old datasets (around 2010) discussed |
| [13] | Both | ✗ | ✓ | ✗ | Lack analysis of IoT-related datasets |
| Our | Both | ✓ | ✓ | ✓ | With ML, DL, and non-ML detection schemes, and analysis impact of APT attacks in the IoT context for the first time |

## 2 SECURITY IN IOT

It is widely recognized that IoT technologies and applications are still in their infancy [42]. Several challenges restrain the development and application of IoT, including technical limitations, lack of standardization, and limited security and privacy components [85]. The acceptance and ubiquity of new IoT technologies and services largely depend on the status of data security and privacy protection, which are two challenging issues for IoT because of its deployment, mobility, and complexity [43]. The study in [98] describes IoT in a three-layer architecture, namely perception, network, and application layers. Security challenges associated with each layer are shown in Fig.3.

### 2.1 Security Features of IoT

The security requirements of an IoT system vary based on its purpose. An IoT system should perform its defined functions, prevent attacks, and operate with resilience when under attack [161]. A secure and resilient IoT system needs to satisfy confidentiality, integrity, availability, and authentication features.

- **Confidentiality** aims to protect information from being accessed by unauthorized parties. Many entities are integrated into an IoT system, and each device could reveal collected data to other nodes within the network. For instance, the study in [132] demonstrates that IoT-based e-health systems, where confidentiality is a high requirement, are vulnerable to third-party attacks. Consequently, a set of management approaches is required to ensure data security and confidentiality [10].

- **Integrity** refers to ensuring the authenticity of a piece of information (e.g., the information is not altered and the data source is genuine). Data integrity ensures the quality of service and data security and privacy [88]. With the increasing availability of cloud computing, coupled with ever-increasing volumes of data, maintaining the data's integrity is critical, especially for outsourced data or third party data. Measures are taken to ensure integrity, include 1) controlling the physical environment of networked terminals and servers, 2) restricting access to data, and 3) maintaining rigorous authentication practices. For instance, the authors in [140] demonstrate an authorization framework distributing processing costs between constrained IoT devices and less constrained back-end servers while keeping message exchanges with the minimum constrained devices. The study in [71] introduces a number of authentication techniques to address data integrity issues.

Table 3. List of abbreviations

| Abbre | Description | Abbre | Description |
|---|---|---|---|
| 3GPP | 3rd Generation Partnership Project | AML | Adversarial Machine Learning |
| APT | Advanced persistent threat | CI | Critical Infrastructure |
| CNN | Convolutional Neural Networks | CPPS | Cyber-Physical Production System (CPPS) |
| DAE | Deep AutoEncoder | DBN | Deep Belief Networks |
| DBSCAN | Density-Based Spatial Clustering of Applications with Noise | DL | Deep Learning |
| DMZ | Demilitarized Zone | DoS | Denial-of-Service |
| DP | Data Type Probing | DS2OS | Distributed Smart Space Orchestration System |
| DT | Decision tree | FCN | Fully Connected Network |
| HMI | Human-Machine Interface | ICS | Industrial Control System |
| IDS | intrusion detection system | IIoT | Industrial IoT |
| IoT | Internet of Things | KNN | K-Nearest Neighbor |
| LR | Linear Regression | LRDoS | Low-Rate One-Class DoS |
| LSTM | Long Short-Term Memory | MC | Malicious Control |
| ML | Machine Learning | MO | Malicious Operation |
| OT | Operational Technology | PASTA | Process for Attack Simulation and Threat Analysis |
| PLC | Programmable Logic Controllers | RNN | Recurrent Neural Network |
| UML | Unified Modeling Language | WS | Wrong Setup |

• **Availability** is the ability to access data or services when and where needed [131]. The priority of IoT security is to ensure the availability and functionality of each IoT application [171]. The study in [188] demonstrates a concept based on probability theory for measuring the reliability and availability of the IoT devices in the network.

• **Authentication** is the process of verifying a person's identity or device and preventing unauthorized access, and misuse of related personal information [149]. In an IoT context, it is essential to deploy an authentication process because of the nature of the IoT. Specifically, several entities (e.g., equipment, users, service providers, and processing units) participate in an IoT system [180]. The authors in [91] introduce an access control policy for authentication based on the device's particular role(s) in the associated IoT network.

## 2.2 Security in Industrial IoT

An Industrial Control System (ICS) is a general term for a series of systems used for controlling industrial processes [178]. They are widely applied in power generation, petroleum extraction and processing, water conservancy, and other industrial fields [95]. The creation of the IoT has allowed for new opportunities in industrial development

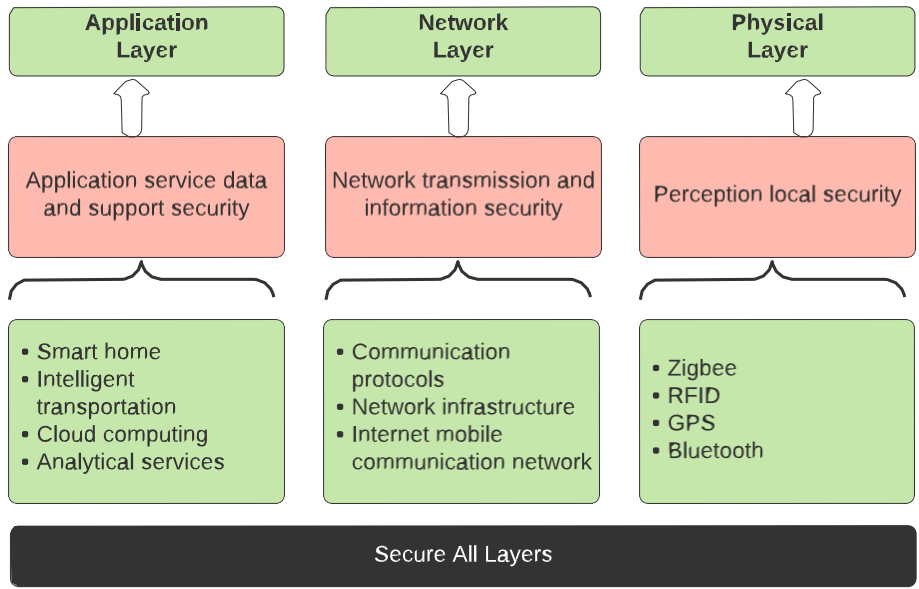

Fig. 3. IoT security challenges in each layer [98]

[189]. For example, various sensor networks provide the necessary supervision and decision data for an ICS, which in turn allows the ICS to control the production process more efficiently and accurately [39]. However, wireless communication and vulnerabilities in IoT expose ICS to internal security risks [18]. We now introduce the essential ICS structures and analyze their vulnerabilities.

IIoT systems are targeted by cybercriminals because the content accessible in those networks can include valuable and confidential manufacturing, and production data [87]. It is also vital to protect IIoT systems against attacks because such systems are often interconnected to other information and communications technologies among the manufacturing processes. Internet-enabled IoT has allowed these networks to become part of the CI for business operations, including industrial control systems, which is a general term for a variety of control systems [24]. The studies in [29] present CI in Canada, which has ten categories of CI, including those mentioned for the US as well as others such as health, food, and water.

Enterprises protect their ICSs via firewalls, which restrict and supervise the communication between the internal and external networks, aiming to ensure the overall security of the network systems [111]. According to the location and function of firewalls, the structure of ICS can be divided into three types (e.g., dual firewalls, single firewall, and multi-firewall facilities) [7]. Properly configured firewalls can protect an ICS, use of Intrusion Detection Systems (IDSs), and controls on application-level privileges [7].

ICS LAN is equivalent to a Demilitarized Zone (DMZ) for the business LAN in single firewall architecture. The purpose of a DMZ is to add an additional layer where an external network node can only access public resources permitted in the DMZ while the rest of the network is firewalled. Dual firewalls are the most commonly used protection structure, where two firewalls separate Internet, business LAN, and ICS LAN, as shown in Fig. 4. Generally, the firewall between business LAN and the Internet is operated by IT staff, while the other firewall is operated by ICS staff. The dual firewall structure offers a higher security level by denying unauthorized access from external networks than using only one firewall. Business and ICS staff jointly control the single firewall that is utilized for protecting ICS from cyberattacks. Finally, multi-firewall infrastructures are suitable for large-scale distributed control systems. This structure is based on a dual firewall containing multiple ICS LANs connected with the business LAN. Each connection between the ICS LAN and the Business LAN is equipped with a firewall. When communication is required between ICS LANs, some business LANs will be selected as intermediate paths.

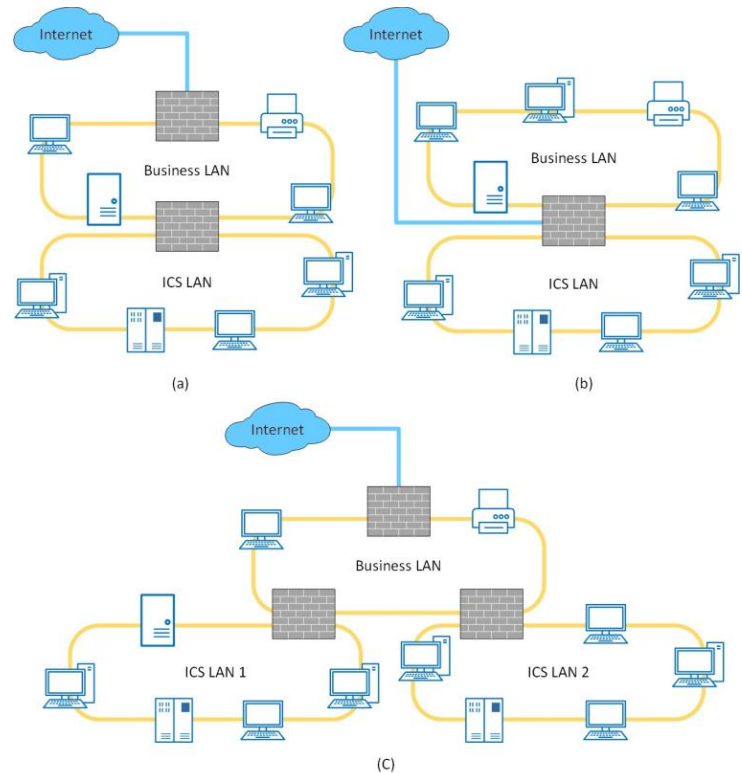

Fig. 4. Firewall architectures. (a) single firewall architecture, (b)dual firewall architecture, (c) multi-firewall architecture.

## 3 COMMON ATTACKS, APT ATTACKS, AND THREAT MODEL ANALYSIS IN IOT

The literature in this area is typically driven by the availability of datasets. Various datasets have been used in many of the papers examined in this article, consequently leading to different standards to classify attack types.

### 3.1 Common Attacks in Industrial IoT

IIoT, as a derivation of IoT, experience similar types of cyberattacks which is vulnerable to various range of attacks. Networks, such as IIoT, are an essential asset in a business [125]. Adversaries take advantage of cybersecurity weaknesses in IIoT to 1) obtain confidential information, 2) disrupt operations, and 3) cause damage potentially to the reputation and interests of an enterprise under attack. CI is a prominent part of enterprises and industries. For example, in 2019, a CI was attacked every 14 seconds on average [5]. Meanwhile, the work in [5] predicts the losses caused by cyberattacks targeting enterprises to be as high as $11.5 billion at the beginning of 2022. Addressing ICS vulnerabilities in CI will minimize the negative financial impact while creating an opportunity and need for future cybersecurity research.

Among the ICS vulnerabilities [7], the simplest one is that an attacker can directly send commands to data acquisition equipment. Since most Programmable Logic Controllers (PLCs) and data acquisition servers lack basic authentication, they generally accept any information sent to them. All of the adversaries aim to establish a connection with the device via breaking in through the firewall and then send appropriate commands to take control.

Another effective attack method for the adversaries is to export the operator's Human-Machine Interface (HMI) directly to an external device. Off-the-shelf tools can complete this operation under Windows and UNIX environments. Finally, the adversary can use a Man-in-the-Middle attack in which the attacker plays the role of a middleman. If the attacker needs to know the ICS protocol, they can change data packets between the HMI and the server to both deceive the operator and take control of the ICS [46].

Below is a list of some of the attacks in the IIoT literature:

- **Zero-day Attacks**: Zero-day exploits refer to the time it takes from the detection of an attack to fix the design flaws that developers have not recognized the time of release and until the vulnerability is overcome [27]. Such attacks may occur on field devices and servers. For example, using overflowing buffers, an adversary can inject malicious executable code into running programs in order to take command of various industrial processes.

- **Non-prioritization of tasks**: Non-prioritization of tasks is a severe flaw in many types of real-time ICS [22]. As an example, memory sharing among the tasks with equal priority results in serious security issues. Accessibility to create an Object Entry Point in the kernel domain is a feature of VxWorks, a proprietary real-time operating system, resulting in loopholes.

- **Database Injection**: Database injection exploits the vulnerabilities in an ICS system [26], which widely occur in SQL-based databases. The hackers send fake SQL commands from a web server to a database server, for example, via a supervisory control and data acquisition system, subsequently obtaining unwarranted access to both data and digitally controlled industrial processes.

- **Vulnerabilities in legacy systems**: In association with a legacy system, ICS usually lacks sufficient user and system authorization, and data authorization verification [40], leading to attackers gaining unauthorized access to the system. When faced with a new generation of ICSs, older SCADA controllers and the associated communication protocols cannot be encrypted. Attackers can discover user names and passwords by sniffing software that allows the user to monitor ("sniff") users' internet traffic in real-time, capturing all the data flowing to and from the user's computer.

- **Default Setting**: Through enumeration methods, an attacker can easily crack unsafe devices with default passwords and settings [6], and then control the ICS and implement a corresponding network attack.

- **Remote Access Policy**: When a SCADA system is connected to an unaudited dial-up line or a remote login server, an attacker can easily access the business LAN, and ICS LAN inside the enterprise through the back door [7].

- **DDoS Attacks**: Invalid sources and limited access control allow attackers to destroy Operational Technology (OT) systems by performing distributed DoS attacks on vulnerable unpatched systems [135]. DDoS attacks can be precisely divided into two forms: bandwidth consumption type and resource consumption type. They all require a significant amount of network and equipment resources through many legal or forged requests to achieve the purpose of paralyzing the network and the system.

- **Web Application Attacks**: Traditional OT systems, including HMIs and PLCs, are increasingly connected to networks and accessible anywhere via a web-interface. As a result, unprotected systems are vulnerable to cross-site scripting and SQL injection attacks [128]. Web application attacks are popular APT attacks, that are widely used by cybercriminals.

## 3.2  Introduction to Advanced Persistent Threats

An APT aims to exfiltrate information and control equipment via obtaining access to victims nodes over a long time window [54]. An APT is challenging to mitigate due to combining multiple techniques, which demonstrate expansion in the quantity of IoT-based applications within the development of IoT [151]. Botnet of APT has become one of the most critical attacks in IoT networks, which utilizes malware to take control of victim's equipment connected to web services [167]. The computer dominated by an attack system is a botmaster, where the equipment under control of the botmaster is a bot. Botmaster can instruct a single bot simultaneously and work cooperatively for an individual task completion. With various bots under control of a botmaster, the botmaster is able to inject a large-scale activity in an IoT system, which is more harmful than a conventional malware attack [173]. Botnet attack life cycle can be summarized as spreading/injection, control and application [57]. An APT could remain disguised within the target system to avoid being detected which gains paramount attention in recent days [33]. For instance, the study in [168] demonstrates a framework for botnet attack detection via investigating DNS services in IoT of smart cities scenarios, which integrates deep learning algorithms to inspect DNS logs with motivation to reduce false alarm rate. In [151], the authors implement a deep learning-based approach to identify botnet attacks which aggregates network data in IoT systems and extracts features among them such as connection log data. The proposed system contributes a higher detection performance than normal ML algorithms. Moreover, the authors in [127] investigate a detection system for botnet attack-based domain generation algorithms and DNS homographs detection with deep learning models. Zero-day exploits combined with social engineering techniques result in almost undetectable behaviour [155]. It means APTs lead to critical damage and substantial financial loss due to an extended period and response costs to be able to identify them [101]. An APT attack primarily contains the following three characteristics.

**1. Advanced**: Attackers do not follow an immutable attack pattern but analyze the target network so to formulate the most effective attack strategy exploiting zero-day vulnerabilities, phishing emails, and SQL injections and several others [16]. In addition to computer attacks, social engineering and psychology are also used to guide target network users into traps unknowingly and become links in the attack chain [68].

**2. Persistent**: From the beginning of the attack until the attacker obtains the desired information, the duration of an APT attack is usually in months or even years [175]. Hackers have enough time to explore the victim network and adjust the attack strategy according to the actual situation in order to hide their traces better. Simultaneously, compared to a large amount of normal traffic, the data from the attacker is insignificant, which makes it difficult for network administrators and IDSs to detect [154].

**3. Threats**: APT attacks are complex and require the initiators to master many attack methods and computational theories so it is difficult for individuals to complete such complex tasks. The funded hacker organizations are orderly and usually have the will of the organization or the country behind them [16]. After the target information is stolen, the hacker carries out follow-up work such as clearing traces, making it difficult to carry out the tracking investigation work, and the company or the country suffers enormous losses [121].

From the above characteristics, it is clear that the APT attack integrates the currently known attack methods, and according to the customized attack strategy, it infringes the target network for a long time, which is highly secretive and harmful. The six primary processes of an APT attack are shown in Fig. 5. The yellow color blocks (e.g., Reconnaissance, Initial Compromise, Post Stage) indicate the initial and final stages of the APT attack, which mainly include information collection and trace spiritual work. The red color blocks (e.g., Lateral Movement, Pivoting, Data Exfiltration) represent the attack stages of APT, in which hackers gradually conquer the network domain and leak data. Hackers may change the sequence of the three attack stages according to the network environment.

**Reconnaissance**: In the Reconnaissance stage, the adversaries aim to collect enough information to prepare for the APT attack and develop appropriate attack strategies for the target network. The information collected social engineering information and target network's infrastructure information. Adversaries collect social information, hobbies, and social networking of users of the target network [187]. This information is used to assist in completing the *Initial Compromise* phase. Target network's infrastructure information includes the models of network switches, routers, and the communication protocols used in the network [117]. Attackers use covert attack methods based on this information to ensure their security and construct a reasonable attack path to bypass the network administrators and IDS.

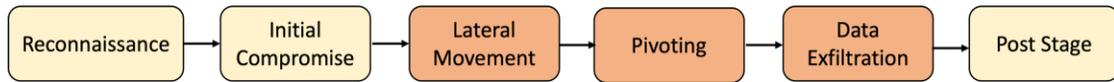

Fig. 5. APT Attack's 6 stages

**Initial Compromise**: In this stage, the attacker uses the information collected in the *Reconnaissance* stage to attack the network user or the target network's vulnerabilities to obtain the administrator privileges of the compromised computer. Then, they use this node to carry out large-scale attacks on the target network. *Initial Compromise* mainly categorizes two forms. The first form is mainly for computer users [38]. After collecting enough social engineering information, adversaries attack network users through different techniques such as phishing emails or links to trick users into clicking in order for the attackers to gain administrator privileges. The second type aims at computer vulnerabilities. By understanding the computer's operating system, patch version, and installed software, attackers use known vulnerabilities, and advanced hackers even use zero-day vulnerabilities to obtain root privileges [110].

**Lateral Movement**: In this stage, the APT attackers use the compromised node to obtain other computers' credentials in the same network domain. Attackers use various methods to move laterally according to the actual situation in the network domain [158]. Commonly used techniques include pass-the-hash, pass-the-ticket, and Remote Desktop Protocol [16]. It is worth mentioning that lateral movement and reconnaissance within the network domain can occur simultaneously. Unlike the original reconnaissance, *Internal Reconnaissance* mainly refers to the hacker's activities inside the target network. After obtaining the administrator credentials, the adversaries obtain essential information such as the computer's running process, IP, port numbers. With the deepening of reconnaissance, attackers can also modify the attack strategy at any time, making APT attacks more difficult to detect.

**Pivoting**: For companies and organizations, a multi-location office has become a typical application scenario. Therefore, the internal network is often composed of two or more remote subnets responsible for different tasks. Each subnet has a domain controller and several user computers. The domain controllers are connected by a communication channel such as a VPN. When an APT attacker encounters the need to move laterally across different subnets, pivoting attacks become the most powerful tool [19].

**Date Exfiltration**: This stage mainly uses the session established by the previous attack for information and file return. To avoid being noticed by the IDS system and anti-virus software, attackers usually use the tools that have been written to cut the files and then transfer them. Advanced hackers will take the different cut parts to different DNS servers and then collect the sliced file fragments to reduce the possibility of discovery [108].

**Post Stages**: When the hackers obtain all the required information, the APT attack phase is over. The primary purpose of the post stage is to clean up the traces of APT attacks and prevent follow-up investigations [121]. During an APT attack, the attacker installs various malware on the compromised computer. Multiple logins and reconnaissance generate log files and browsing records. Therefore, cleaning up these traces can well hide the APT attack and avoid tracking investigation.

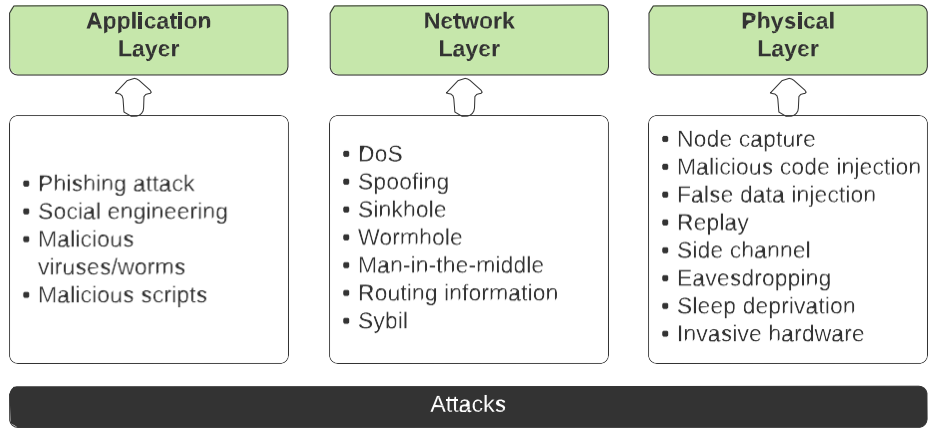

Fig. 6. Main attacks in each layer

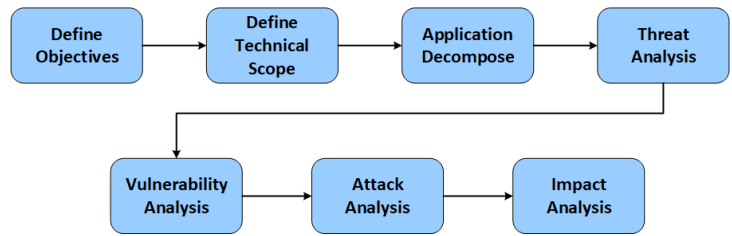

Fig. 7. The PASTA Model

## 3.3 Attack Analysis with Threat Model

Threat modeling is a systematic way to help security experts, software engineers, and organizations determine the assets, structure, and potential vulnerabilities of target systems and prevent implied attacks [160]. IIoT, specifically ICS, confront significant security challenges due to slow patching processes, unsafe protocols, and other security problems, leaving a relatively open window for attackers to exploit vulnerabilities. For example, Industroyer, a malware framework that was discovered in 2016, targeted the electric grid in Ukraine's capital, causing a short-term power outage in that area [1]. Therefore, threat modeling can be a useful process that can optimize network security by defining objectives. Threat modeling is vital for IIoT systems. In this paper, the PASTA model is reviewed in respect to IIoT systems. As shown in Fig. 7, there are seven stages of PASTA, include defining business objectives, defining the technical scope, application decomposition, threat analysis, vulnerable detection, attack enumeration, and impact analysis [144].

Fig. 8 shows the high-level architecture of an ICS. As noted, ICS contains four principal components that are control and sensing systems, data services, asset management, and the institution's private network(s) [102]. Control and sensing systems utilize on-site hardware to collect data from the environment and operate the overall network [84]. This system is similar to the Perception layer in Fig. 3. The data collected from the control and sensing systems will be stored and managed. Asset management handles and displays the data collected by sensors and actuators that form links with the organizations' private network. The network devices and typologies in Sec. 2 with devices (e.g., network cameras, PLC, RTU, and HMI) are connected to the ICS LAN, and the business LAN. Various protocols (e.g., S7, Modbus, Distributed Network Protocol 3 (DNP3), EtherIP, HTTP, FTP, SSH, Telnet, and UPnP) can be used for the sensing component of the control system. The application layer utilizes data services such as SQL, SSH, and HTTP. However, these services can be exploited by attackers. Moreover, control devices like PLCs may use mainstream operating systems like Windows or Linux, thereby increasing the threat surface and the vulnerabilities to the level of an entire enterprise's digitally controlled operating environment [4].

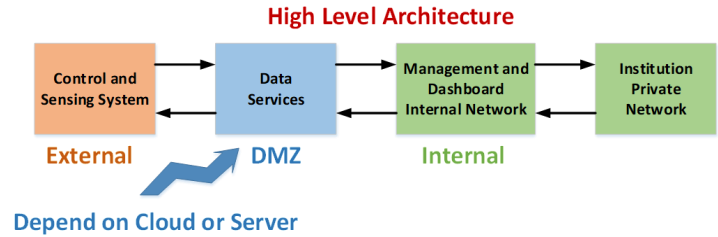

Fig. 8. High Level Architecture of ICS

ICS decomposition is shown in Fig. 9, which illustrates more detailed structures such as the hardware, software, and entry points. Control systems are comprised of multiple hardware components, including PLCs, RTUs, SCADA devices, or distributed control system devices. Data within control systems appear as raw ICS-based data (e.g., serial communication), ICS-based network traffic (e.g., DNP3, Modbus, and EtherIP), and traditional network traffic (e.g., TCP, UDP, ICMP, and ARP) [102]. The data flows from these networks into the management component of the current control system. Several services are deployed in the control system to handle the raw data from control systems, such as traditional and SCADA servers.

Either the raw data from control systems or the processed data from control system management are fed into a remote data service, which may use SQL or cloud techniques. The structured data are transferred from the network's collection point to the management process, where cybersecurity is applied and further displayed on the dashboard in the HMI systems. By analyzing the data flow in ICS, assets can be identified as hardware (e.g., PLCs, RTUs, or network cameras) or software such as multiple services leveraged to manage the control systems and data collected from the control systems. Moreover, the entry points contain the control system devices, internal networks, and data servers, amongst which IoT devices are the most vulnerable components due to the lack of security protection compared to servers or PCs. The ICS system will also be connected to internal private networks, resulting in the potential for a broader threat surface. Nonetheless, as our technical scope is limited to external attacks on IIoT, this work will not focus on industrial networks' internal attacks.

Threat analysis assesses the possible threats targeting the assets and identifies the most probable attack scenarios. As a consequence, ICS and SCADA systems are high-value targets for attackers. For example, in 2019, the power sector received 6% of all ICS and SCADA attacks and incidents in the top ten industries [6]. As pointed out by IBM in 2019 [3], attacks targeting Windows Server Message Block protocol can be automatically implemented, making it easier for low-level attackers to compromise devices. Commodity downloaders, such as Emotet and TrickBot, are often leveraged by attackers to execute the malware on victims' machines. Meanwhile, PowerShell is frequently used to download attack tools and native functions, such as PSExec or Windows Management Instrumentation, which are also exploited to help malicious users move laterally, which can be hard to detect.

A significant number of vulnerabilities exploit improper authentication or excessive privileges. We mean too much privilege assigned so we may keep 'excessive privilege', among which 64% [6] can be utilized remotely by attackers. These vulnerabilities involve permissions, privileges and access controls, improper input validation, improper restriction of operations within the bounds of memory buffers, path traversal, information exposure, improper access control, command injection, SQL injection, scripting, and improper authentication. Besides, 82% of these vulnerabilities exploit networks attack vectors; hence network equipment is the crucial component in delivering commands among ICS devices [6]. This equipment vulnerability offers ICS attack surfaces consisting of personal computers on internal networks, IoT devices, control system devices, HMIs, and databases.

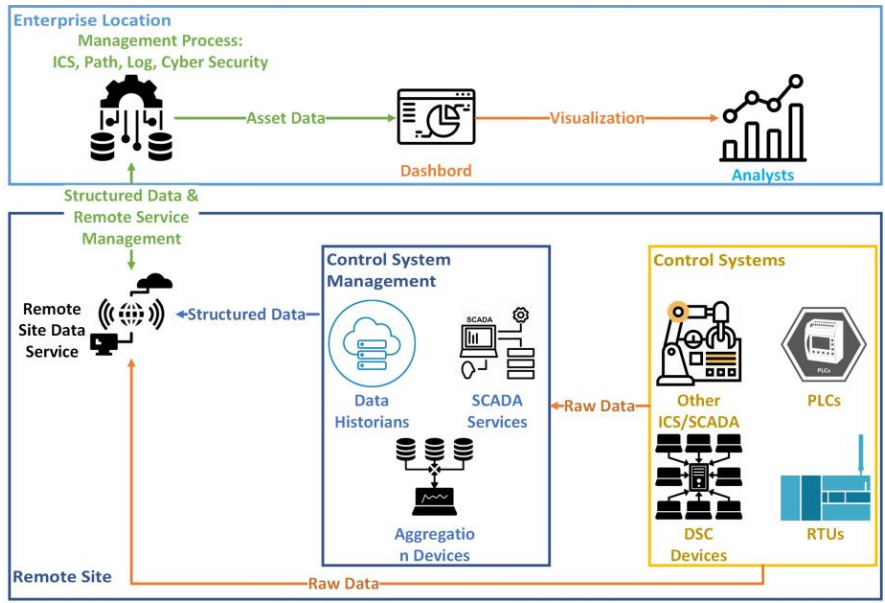

Fig. 9. ICS decomposition

## 4 INTRUSION DETECTION APPROACHES IN IOT

This section provides a review of IDS for IoT, emphasizing ML approaches. IDSs can be categorized into several types based on the detection mechanism, namely anomaly-based, signature-based, hybrid, collaboration, and specification [182]. The anomaly-based, signature-based, hybrid IDS, and collaborative methods are introduced in the subsequent sections. Specification-based approaches require a collection of rules and thresholds to determine network component behaviors. When network behavior deviates from specification definitions, specification-based approaches kick in to detect intrusions [182]. The survey in [92] points out that a specification-based IDS in a smart grid can achieve superior performance. However, only a few papers investigate the integration of specification-based approaches and MLs. Therefore, we do not focus on specification-based IDS in this review. Table 4 summaries the different approaches to implement network intrusion systems [141], (e.g., statistical, data mining, and ML), among which ML based methods [153], are used in hybrid detection systems [61].

### 4.1 Signature-based Methods

A signature detection system differentiates malicious from standard traffic patterns or application data [69]. If the network behavior matches a stored pattern or signature of nefarious code, it is marked as abnormal. Thus, signature detection can identify attacks but only if a signature (e.g., label) for the attack is available. Hence, the method performs poorly in cases representing a new kind of attack [182].

The minimal computing power and memory limitations of IoT devices combine to create a security challenge compared to traditional network and computer security systems. The study in [113] shows a comparison between packet payloads and attack signatures systems while considering the difference in the computational cost of the two detection systems. Consequently, the authors demonstrate a lightweight security system using a pattern-matching engine to detect malicious code. The engine runs on resource limited equipment by restraining memory usage of a simulated IoT system. Meanwhile, they present two innovative techniques called 'auxiliary shifting' and 'early decision' with motivation to alleviate degradation of performance given the constraint on sensor-level computational power and available memory. Based on these proposed techniques, the number of matching operations on resource-constrained systems, such as an IoT network, decreases, resulting in enhanced efficiency under constrained conditions.

The authors in [106] demonstrate a lightweight digital signature approach to ensure reliable communication between smart equipment IoT networks with reduced number of operations. The proposed method ensures security via guarding against traffic analysis attacks in IoT systems. A DoS detection architecture is demonstrated for the 6LoWPAN protocol for low-power lossy networks [107].

## 4.2  Anomaly-based Methods

An anomaly detection system compares a captured pattern in a data stream to pre-stored baseline patterns of known threats. If the patterns match, action is taken accordingly the operational needs of the time. However, the downside of anomaly detection is the potential for inevitable false alarms. Based on applied techniques, anomaly-based intrusion detection methods can be categorized according to their use of learning, statistics, and miscellaneous techniques [156]. In a statistics-based approach, network traffic activities capture and create a profile representing the random behavior of network traffic. This profile takes various essential features into account (e.g., traffic rate, number of packets, connections, multiple IP addresses). The anomaly detection procedure for network traffic utilizes two datasets;

One contains the current observed profile, and the other is a previously trained statistical profile. As network events occur, the existing profile is determined, and a quantitative value is calculated by comparing the two actions. Thus, IDSs can figure out anomalies based on the calculated scores. On the one hand, the statistics-based methods lead to increased detection accuracy. However, the downside of statistical methods is that malicious users could hack into the system by training the system and its parameters by injecting malicious code.

The category of miscellaneous detection techniques consists of all the approaches not included in the statistical and ML groups and includes using such approaches as control-flow graphs [169], finite-automata [30], and description languages [176].

For instance, the study in [67] provides a framework to determine intrusion based on the general-purpose Unified Modeling Language (UML). The authors design a UML profile called UMLintr that allows software developers to specify intrusions using UML notations extended to suit the context of various intrusion scenarios. The framework uses the expressiveness of UML and eradicates the requirement of using attack languages. It is unnecessary to study the UML language for attack descriptions, thus reducing developers' efforts in specifying intrusion scenarios. The authors in [176] introduce a description language (e.g., n-grams) for intrusion detection. They show both anomaly detection and classification based on n-grams. Besides, they develop criteria for utilizing data using different schemes and applying the defined criteria to web intrusion detection. Although miscellaneous detection techniques are flexible and scalable, they are not effective for high quality and high volumes of data due to time-consuming data management caused by a low-power and limited memory environment [52]. Lately, a number of works deploy anomaly techniques to defend against attacks in an IoT system, including wormhole attack, sinkhole attack, DoS and many others. For instance, the study in [45] investigates an anomaly-technique based framework for wormhole attack detection, which is rarely discussed in recent literature. The authors demonstrate wormhole attack at 6LoWPAN adaption layer of RPL network that is mitigated by an implemented IDS in Contiki OS. The authors in [94] introduce PRDSA scheme based on probe route to prevent sinkhole attack and ensure security for IoT systems. The PRDSA framework can be utilized for detecting, bypassing and locating sinkhole attack simultaneously. Furthermore, an IDS system built on anomaly technique is studied in [97] to address DoS attack in IoT. IoT systems are vulnerable due to DoS attacks which are attacked by malicious users to deliver a large number of valid data. Thus, the authors propose a selective authentication-based geographic opportunistic routing framework to resist DoS attacks in IoT which performs about 50% of computational cost and provides a reliable data distribution.

In recent years, ML algorithms have emerged as a tool to assist in anomaly-based IDSs. ML techniques are used to build explicit or implicit models that utilize pattern analysis to classify data. A ML-based network intrusion system can change its execution strategy as it obtains new information [52]. There are a number of ML algorithms used in anomaly-based detection systems including SVM, Bayesian, neural network, clustering and outlier detection [156] [52]. The study in [56] demonstrates a comprehensive review introducing ML-based schemes targeting IoT for smart healthcare. The main disadvantage is the need for relatively expensive computation resources to perform the necessary data processing.

### 4.3 Hybrid Methods

As for a hybrid detection system, it combines the signature-based and anomaly-based approaches to overcome their drawbacks and benefit from their advantages [147]. Hence, hybrid detection systems can synthesize the benefits of two methods such as the detection of "zero-day" attacks and the detection of signatures of known attacks. Traditional IDS is difficult to be deployed straightforwardly in IoT networks because of various communication schemes, protocols, technologies, and particular services. Meanwhile, hybrid-based IDS is able to adapt techniques and services in real-time which is a serious issue in conventional IDS [163]. The authors in [163] design a hybrid IDS based on timed automate controller approach to detect DoS attack, control hijacking attack, zero-day attack, and replay attack in IoT environments with promising accuracy (e.g., 99.06%). The study in [147] introduces hybrid techniques namely using Gaussian Mixture clustering with Random Forest Classifiers as well as using K-means clustering with Random Forest Classifiers in order to detect intrusions. The authors report that this proposed hybrid approach decreased the false alarm rate to 0.04%. The authors in [31] demonstrate a hybrid IDS, called INTI, which was designed to detect sinkhole attacks in the routing services within an IoT network. INTI links anomaly-based concepts and specification-based methods to monitor and audit packets between nodes and extracts reputation and trust features for node evaluation.

### 4.4 Collaborative Methods

Collaborative IDSs attract attention for IoT applications. It differs from the other three types of three methods. Collaborative methods do not need a high demand for resources and reduce delay in identifying suspicious activities since they monitor and collect abnormal acts. Collaboration between IoT equipment allows obtaining knowledge from various host and network equipment to result in promising detection performance [162]. The study in [21] demonstrates a collaborative IDS, namely COLIDE, to identify attacks in IoT networks, which allows an efficient detection with low latency via collaboration among sensors and devices with limited resources in IoT systems. The authors in [83] present a collaborative IDS to boost detection accuracy, which integrates a semi-supervised ML model to address the missing label problem. It is possible to observe unlabeled network samples due to the expensive cost of labeling numerous instances in a network. The upside of collaborative IDS is reported to be lower false alarm rate and efficient performance with unlabeled samples.

Table 4. IDS Techniques

| IDS Tech | Ref | Attack | Benefits | Open Issues |
|----------|-----|--------|----------|-------------|
| Signature | [69] | DoS | Implement 2 types of DoS | Low performance |
| | [106] | Traffic Analysis | Low computing time | Real-time attack detection |
| Anomaly | [45] | Wormhole | Wormhole rarely discussed | Low positive detecting |
| | [94] | Sink hole | Promising performance | Identify attack earlier |
| | [97] | DoS | 50% of the computational costs | Long delay |
| Hybrid | [31] | Sink hole | Low false prediction | More types of attack detection |
| | [163] | DoS | 99.06% detection accuracy | Time consuming |
| Collaborative | [83] | DoS, Probe, R2L U2R | Low false alarm; efficient | Vulnerability to various data characteristics |
| | [21] | Distributed attacks | Low workload | Require of RAM increase |

## 5 MACHINE LEARNING IN IOT SECURITY

ML can be utilized to implement a data-driven intelligent attack detection platform for IoT systems [136]. In this section, we study various ML-based approaches that are used to detect network intrusions in IoT applications and categorize them as superv ML, deep learning, and unsupervised ML algorithms as shown in Table 5, Table 6, and Table 7, respectively. Among the papers considered in this work, supervised ML methods are widely used to guarantee accuracy and efficiency; whereas, unsupervised methods are not as commonly used for intrusion detection in IoT networks. Meanwhile, deep learning approaches confronts issues such as higher requirements of computation resources and longer response time for prediction when compared with ML algorithms.

### 5.1 Supervised Machine Learning Methods For Network Intrusion Detection

Adaboost stands for Adaptive boosting, the first practical boosting algorithm, which is used widely in various applications [137, 165]. The study in [123] introduces an IDS based on the Adaboost model for DoS attack detection. The work applies a public IoT dataset MQTTset for evaluation that achieved up to 95.84% overall accuracy. Meanwhile, the authors in [170] present a framework with Adaboost for Low-Rate One-Class DoS (LRDoS) attack detection, that is more difficult to identify when compared to conventional DoS attack. Moreover, the proposed approach is adjusted according to sample weights, which tackles imbalanced dataset issues. Under NS2 simulation-based test environment, the presented method performs 97.06% detection rate for the LRDoS attack.

Decision Tree (DT) algorithms depend on some pre-conditions conditions and select a univariate split for the root of the tree and recursively repeats the process [134]. Pruning reduces the size of trees and is performed after a complete tree has been built. The study in [142] shows that DT-based detection framework achieving up to 99.99% accuracy to identify Bot-IoT attacks using the public dataset Bot-IoT. Furthermore, the study in [72] demonstrates 95.64% overall accuracy in detecting APT attacks in NSL-KDD dataset via DT algorithm which deploys an early detection of these APT attacks. Thus, the proposed approach is suitable for APT attack detection due to APT long time presence in compromised systems.

Table 5. Summary of machine learning algorithms In IoT

| ML | Ref | Attack | IoT Dataset | Benefits | Open Issues |
|---|---|---|---|---|---|
| Adaboost | [123] | DoS | MQTTset | High accuracy | Apply other ensemble models |
| Adaboost | [170] | LRDoS | Private | Optimize Adaboost | Evaluate on real networks |
| DT | [142] | Botnet | Bot-IoT | 99.99% accuracy | Apply to other researching |
| KNN | [81] | DoS, DP, MC, MO, SC, Spying, WS | DS2OS | 99.99% detection rate | Apply deep learning |
| LR | [63] | DoS, DP, MC, MO, SC, Spying, WS | DS2OS | High accuracy | Build a robust algorithm |
| LR | [130] | Brute force, Broker Spoofing and Malformed MQTT Packets | Private | Adapt to new schemes | Define more event patterns |
| LR | [44] | Bot | CUT-13 | Detect unknown attack | Long time convergence |
| RF | [63] | DoS, DP, MC, MO, SC, Spying, WS | DS2OS | High accuracy | Build a robust detection algorithm |
| RF | [17] | Blockhole and DoS | Private | Cross network layers | Improve detection performance |
| RF | [63] | DoS, DP, MC, MO, SC, Spying, WS | DS2OS | High accuracy | Build a robust detection algorithm |
| SVM | [63] | DoS, DP, MC, MO, SC, Spying, WS | DS2OS | High accuracy | Focus on real-time data |
| SVM | [109] | EAVESDROPPING, DoS, Spoofing, Man-in-middle | Private | 7ms detection time | Overhead of system collaboration |
| SVM | [70] | BASHLITE and Mirai | N-BaIoT | Metric reflecting security analysis | N/A |
| XGBoost | [172] | vulnerability exploiting, malware infection, abnormal operation, and memory leak | Private | Test under real IoT system | Improve diversity of generated DT |

A K-Nearest Neighbor (KNN) algorithm is a data classification method that estimates the likelihood that a data point will become a member of one group or another, depending on which group the nearest data point belongs. It means a KNN classifier is a distance function, in effect an application of the Pythagorean theorem, which measures the difference or similarity between two instances [11]. The KNN ML approach is practical in several network categories (e.g., IoT, industry WLAN IoT, and self-designed networks) where each network could be affected by different attacks. From the literature [100], KNN has been deployed to detect several network attacks such as APT, remote tripping command injection, relay setting change, data injection, ransomware, and DoS.

However, the KNN-based classification approach almost cannot detect R2L and U2R attacks since the number of these attack traffic are extremely low. Therefore, considering the promising performance of KNN for use in IoT networks, various papers have proposed approaches based on this algorithm with the intent to mitigate network attacks [112].

The study in [100] introduces several ML algorithms implementation, including KNN model MLP, DT, NB, and SVM, to detect IDS in IoT systems. However, none-IoT specific datasets are used to evaluate test performance in this literature. The authors in [114] demonstrate a review of AI techniques that ensure security for IoT systems in medical settings, which indicates KNN is one of the leading algorithms according to numerical results. In [81], the authors design a distributed ensemble framework to identify intrusions via fog computing technique to preserve IoT networks. It is worth noting an IoT specific dataset DS2OS is used to evaluate the proposed system performance which achieves up to 99.99% detection rate.

The Linear Regression (LR) algorithm attempts to model the relationship between two variables by fitting a linear equation to observed data. Thus, One variable is considered an explanatory variable whereas the other is a dependent variable [105]. LR demonstrates 100% accuracy to identify network attacks, which is applicable to various types of attacks such as DoS, probing, and Malicious Control (MC) [63]. The study in [130] integrates ML algorithms (e.g., LR, support vector regression) and complex event processing techniques to identify various kinds of intrusions in a real-time healthcare IoT system. Moreover, the proposed approach demonstrates high performance for TCP, UDP and Xmas port scans, and DoS attack detection, with up to 100% accuracy. Roldán et al utilize a private dataset which was assembled real-time data in a designed IoT network.

RF is an ensemble ML algorithm that consists of a number of decision trees [120]. Each decision tree gives a prediction result and an overall result via voting on by all decision trees. For instance, it is assumed that three decision trees are applied in RF for binary classification, class $A$ and class $B$. Two of the trees give prediction results at a point belonging to class A, while one of three trees estimates the sample belongs in class $A$. Then RF considers the data as part of class $B$ since the majority of members voted for class $A$. Multiple class classification follows in the same manner as binary classification. The RF algorithm is deployed to detect network attacks in Ethernet IoT and self-designed networks [63, 165]. The study in [17] demonstrates a two-stage detection framework for blackhole and DoS attacks in mobile IoT systems. The authors collect data in a designed network and feed the data into RF and other ML models. The numeric results show an F1 score in the range of 93% to 99.36% in identifying the two attacks. In [112], IoT botnet attack mitigation is investigated integrating feature selection and post-hoc local analysis stages in RF and some other ML algorithms working procedures.

SVM is well-known for its ability to effectively handle high-dimensional datasets, even if the number of dimensions is higher than the number of samples, while still maintaining its effectiveness. Since only a subset of samples, named as support vectors, is used by SVM in its decision function. SVM is widely deployed in IoT networks to ensure systems security and mitigate attacks. For instance, the study in [62] utilizes several ML algorithms including SVM to detect DoS attack, Data Type Probing (DP), MC, Malicious Operation (MO), scan, spying, and Wrong Setup (WS). The authors in [109] introduce a novel framework SeArch to represent a collective NIDS system in a cloud IoT network which integrates SVM algorithm and other ML/deep learning models to mitigate the threats (e.g., eavesdropping, DoS, spoofing, man-in-middle) in IoT networks. The authors in [179] demonstrate one-class SVM approach to detect attacks and protect IoT systems which reduces the requirement of memory and decreases computation time. Therefore, the proposed approach is able to be applied in a real-world IoT system that typically consists of resource constrained devices. The study in [70] shows a lightweight attack detection system adopting SVM algorithm to identify malicious users aiming to introduce suspicious data into an IoT network. SVM-based anomaly detection is summarized in a comprehensive review paper [66].

XGBoost is an ensemble ML algorithm based on decision trees that uses a gradient enhancement framework [34]. The study [165] report on an XGBoost-based classification method to mitigate network attacks with a detection performance of 97%. The authors combine XGBoost and LSTM together to construct a novel framework in order to ensure IoT devices security [172]. It is worth noting that a real IoT system is established for performance evaluation. The proposed approach achieves promising performance for vulnerability exploiting, malware infection, abnormal operation, and memory leak.

Table 6. Summary of deep learning algorithms

| ML | Ref | Attack | IoT Dataset | Benefits | Open Issues |
|---|---|---|---|---|---|
| DAE | [82] | N/A | N-BaIoT | Less features used | Reduce features |
| DBN | [25] | DoS, Overflow, SSH Brute Force, Malware, Cache Poisoning | Private | Detect malicious activity inside web | Improvement for standard |
| FCN | [118] | N/A | Private | Cost effectiveness | Need more analyses |
| FCN | [44] | Bot | CUT-13 | Detect unknown attack | Long time convergence |
| LSTM | [15] | BASHLITE and Mirai | N-BaIoT | Improve performance | Need improvement for Scan and TCP flood |
| LSTM | [89] | Abnormal data stream | Private | Reduce regression error | Need monitoring sensors status |
| RNN | [119] | Botnet | Bot-IoT | Robust and tackle over-fitting issue | Long training and response time |
| RNN | [58] | Malware | Private | 94% accuracy | Need a large DT |

## 5.2 Deep Learning For Network Intrusion Detection

A Deep AutoEncoder (DAE) consists of two symmetric multilayer feedforward networks (e.g., encoder and decoder).

DAE is used to reproduce the input at the output layer via the encoder and decoder. The number of neurons in the output layer is the same as the number of neurons in the input layer [183]. Researchers apply DAE in various fields such as dimension reduction, anomaly detection, and noise filtering [77]. The study in [96] demonstrates a neural network with three layers for intrusion detection that integrates DAE and some other ML algorithms. The authors utilize an improved dataset based on NSL-KDD via extracting samples in different layers (e.g., application layer, network layer, transport layer, and full layer). The highest accuracy is up to 97.83% with the proposed framework that can reduce detection load and improve detection performance. Moreover, a lightweight DAE is introduced in [82] that achieves a high accuracy up to 99% under an IoT specific dataset N-BaIoT.

Deep Belief Networks (DBN) are a class of deep neural network, which uses probability and unsupervised learning to produce output [64]. Each layer in the DBN learns the entire input, which means the DBN works globally and regulates each layer in order. The network consists of a cluster of Restricted Boltzmann machines, with nodes in each layer connected to all the nodes in the previous and the next layer; whereas, no connection exists in the same layer. DBN is modeled after the human brain, making it more capable of recognizing patterns and processing complex information. Therefore, DBN algorithms are widely utilized to build network intrusion systems for IoT applications. The study in [184] introduces a DBN-based network attack detection method combining a flow calculation and deep learning. The authors in [185] present a compact DBN-based framework for intrusion detection which is adjusted for complex IoT-integrated environment. A public network dataset NSL-KDD is used to evaluate the proposed approach that results in high detection performance for each class (e.g., DoS, R2L, Probe, U2R). However, the studies in [184, 185] verify system performance based on common network datasets rather than an IoT specific dataset. The authors in [25] introduce an enhanced DBN model to adopt current cyber-attacks with more serious alarming in IoT networks which is compared with conventional IDS approaches. The study demonstrates the DBN-integrated framework kept checking attacks (e.g., DoS, Overflow, SSH Brute Force, Malware, and Cache Poisoning.) based on DBN model obtained knowledge.

A Fully Connected Network (FCN) contains fully connected layers where all inputs from one layer are connected to each activation unit in the previous layer [86]. Fully connected neural network algorithms have proven effective at network intrusion detection with promising performance under different network types. A DNN model consists of one input layer, two hidden layers, and one output layer. The three layers are fully connected, which means neurons in each layer have connections to all active neurons in the previous layer. Assuming that there are m neurons in layer $l-1$. Fully connected neural network algorithms are extensively researched to detect network attacks [35, 44, 48, 63, 74, 122, 164, 166]. Specifically, the study in [28] proposes a DNN-based approach for intrusion detection in a smart home IoT network that resulted in 99.42% accuracy. The study in [118] demonstrates a use case on a smart factory deploying a DNN-based intrusion detection approach that is a context-sensitive based system. The authors collect data and utilize it for a ML model to train and test. The approach is applied into a real company for a test that has been proven to be a cost-effectiveness IDS. Meanwhile, the dataset is private which is difficult for other researchers to access. The authors in [44] present a two-stage IDS BotChase targeting bot attack which integrates unsupervised ML algorithm K-means in the first stage and supervised algorithms (e.g., LR, DT, SVM, FNN) in the second stage. The Botchase can identify bots relying on various protocols, which achieves a robustness system to detect unseen attacks. However, the Botchase takes long time to converge.

A Recurrent Neural Network (RNN) is a type of neural network that takes sequence data as input and performs recursion in the direction of sequence evolution [181]. It is usually used to process sequence information such as strings, lists, tuples, byte arrays, and range objects. Like other neural networks, it has an input layer, an output layer, and a hidden layer. However, the value of the RNN's hidden layer depends not only on the current input; but also on the hidden layer result of the last input [181]. The authors in [119] introduce a stacked RNN integrated framework to detect botnet attack in smart home IoT networks that aims to identify attacks in an extremely imbalanced dataset Bot-IoT. The performance results show a robust system proposed which addresses an overfitting issue. Moreover, the study in [58] demonstrates a malware attack detection system based on RNN as a classifier that targets ARM-integrated IoT systems. The authors collect samples as a dataset for RNN model training and testing which consists of 280 malware and 271 normal documents. Finally, a promising detection performance is presented in the evaluation section, in terms of 98% accuracy.

Intending to avoid long-term dependency issues, Long Short-Term Memory (LSTM) is a special type of RNN [65]. It is used in classifying, processing, and making predictions based on time series data, noting that there can be lags of unknown duration between important events in a time series. LSTM remembers information for long periods with their typical behavior. The study in [48] demonstrates three deep learning algorithms (e.g., DNN, LSTM-RNN, and DBN) to evaluate the difference of IDS performance, which reduce false alarm rate by 1% to 5% and boosted detection rate by 4% to 6% using two public datasets (e.g., NSL-KDD and CICIDS2017). The authors in [15] demonstrate a CNN and LSTM hybrid framework for botnet attack detection (BASHLITE and Mirai) which tackles the insufficient resource issue for devices in IoT networks. Meanwhile, the proposed approach increases botnet attacks accuracy, up to 90.88% under an IoT-specific dataset N-BaIoT. Moreover, the study in [89] presents a LSTM model for detecting and correcting abnormal stream data in IoT networks which ensures a stable and robust IoT application with real-time observing and correcting function. The authors utilize a private dataset with temperature data stream in it, consisting of 96, 453 samples collected from sensors.

Table 7. Summary of unsupervised machine learning algorithms in IoT

| ML | Ref | Attack | IoT Dataset | Benefits | Open Issues |
|---|---|---|---|---|---|
| DBSCAN | [50] | N/A | Private | Group similar attacks effectively | Improve defense capability |
| K-means | [44] | Bot | CUT-13 | Detect unknown attack | Long time convergence |
| K-means | [145] | Worm | Private | 93% detection rate | N/A |

### 5.3 Unsupervised Machine Learning For Network Intrusion Detection

Density-Based Spatial Clustering of Applications with Noise (DBSCAN) is an unsupervised learning method that is well-known for clustering samples based on their density and identifying outliers located in low-density areas [138]. The studies in [35, 44] illustrated DBSCAN algorithm to detect network intrusion. The study in [55] demonstrates the DBSCAN-integrated model for outlier determination on IoT applications which boosts data quality. Furthermore, the study in [50] introduces classifying attack categories with motivation to address complicated, various source and arranged attacks in IoT networks.

The K-means clustering algorithm is an unsupervised ML Methods, which originates from a vector quantization method in signal processing [14, 59]. The purpose of the K-means clustering is to divide n points into k clusters so that every point belongs to the cluster corresponding to the nearest mean. K-means are usually applied to the fields of vector quantization, cluster analysis, and feature learning. K-means integrated methods are presented in [44] for attack detection. The study in [159] investigates the development of knowledge structure combining blockchain and IoT together. The authors utilize K-means to analyze intra-cluster relations among samples in each cluster. The authors in [145] present three innovative IDSs for IoT networks targeting worm attacks detection, including K-means-based IDS, DT-based IDS, and a hybrid framework consisting K-means and DT. The numeric results show the hybrid approach performs the highest accuracy and reduced false prediction critically. The study in [44] shows a two-stage IDS integrating K-means algorithm in the first stage to ensure detection performance and detection effectiveness.

### 5.4 ML Defense Methods Statistic Results in Attack Detection

A number of literature is reviewed where popular AI methods are applied for network intrusion detection. Sec. 5.1 lists papers have been reviewed for supervised ML model; Sec. 5.2 reviewed literature using deep learning algorithms for security in IoT platforms; Sec. 5.3 presents two unsupervised ML algorithms (e.g., K-means and DBSCAN) for intrusion detection in IoT networks. The statistical results for the ML applied frequency are demonstrated in Fig. 10. There are **10** papers use the FCN-based approach for detection, which is the most widely utilized frequency. Moreover, supervised ML approach are the widely used to address network intrusion, comparing with unsupervised ML algorithms and deep learning algorithms. In respect to unsupervised ML algorithms, there are limited works deploying for attacks detection in IoT networks. This work studies K-means and DBSCAN algorithms for ensuring security in IoT networks. Unsupervised ML algorithms are typically integrated with supervised AI models for intrusion detection in IoT platforms that are not conventionally utilized independently to identify attacks in IoT systems. In terms of detection performance among supervised ML algorithms and deep learning algorithms, ML algorithms perform promising performance (e.g., **99.99%** by DT in [142]) while deep learning algorithms demonstrate a little bit lower performance (e.g., **94%** in [58]) in intrusion detection in IoT systems. Regarding to time complexity and demanding of resources, deep learning algorithms suffer from computational complexity and long training times [44]. Therefore, testing procedure efficiency needs to be considered when using deep learning algorithms [58].

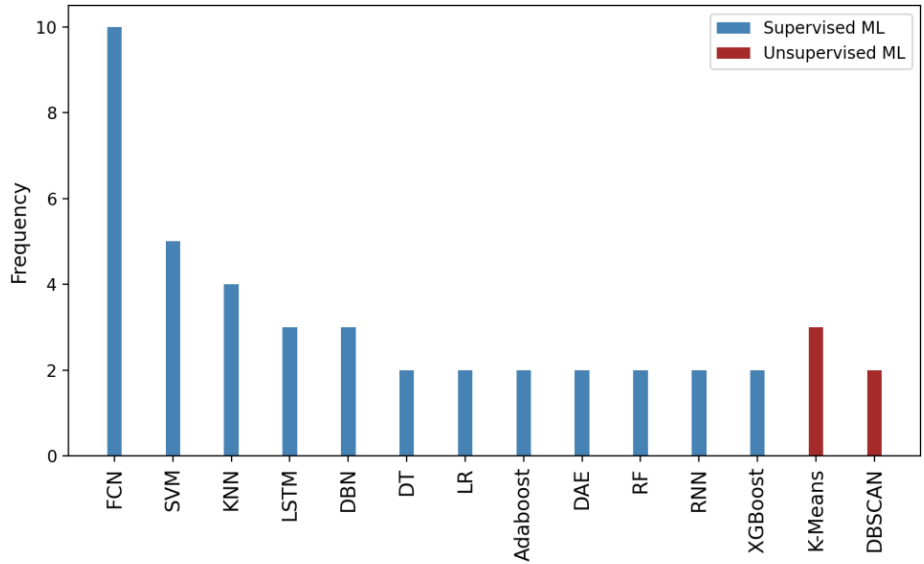

Fig. 10. Machine learning methods utilized frequency of occurrence

## 5.5 IoT Dataset Introduction

IoT-specific dataset is one of most critical roadblocks in IoT security since there are very limited public datasets available for accessing. Meanwhile, it is demanding and time consuming to deploy an IoT network to collect data and generate a private dataset. This section introduces several public IoT-related network intrusion datasets and presents private datasets created and utilized in existing papers. The following contents contain public datasets introduction, attack types and features extracted. Table 8 further summarizes the attributes of IoT intrusion detection datasets.

- **TON-IoT**:
  The TON-IoT is designed for the AI-based IDS among the Industry 4.0 and IIoT environment [8]. It could be utilized to evaluate or validate the efficiency and intrusion detection performance of the AI models. The testbed includes a realistic large-scale network which is deployed by multiple virtual machines and sensors. Numerous attack techniques, including DoS, DDoS, ransomware, against web applications, IoT gateways, and computer systems across the IoT/IIoT network are implemented within the testbed.
- **Bot-IoT**: The Bot-IoT dataset blends real IoT traffic with diverse attack traffic, most notably the Botnet, to address the shortcomings of existing related datasets in terms of collecting total network traffic and attack diversity [80]. The raw traffic in PCAP format and the processed traffic used for ML in CSV format are released to the public for a variety of purposes. Additionally, these files are grouped into subcategories based on the attack types for the convenience of labelling.
- **MQTTset**: The MQTT dataset focuses on intrusions against the MQTT protocol, which is a standard application protocol for IoT networks [123]. The network traffic is gathered from a variety of real-life sensors. The dataset covers five attack types: SlowTe (a novel DoS attack), bruteforce authentication, corrupted data, flooding, and DoS, among which bruteforce authentication can be considered as an APT attack. There are 33 features available, including several TCP head capabilities and MQTT protocol characteristics.

- **CTU-13**: The CTU-13 dataset seeks to collect a large volume of traffic from real-world IoT networks, as well as background and botnet intrusions [53]. Over 90 IoT devices and 13 botnet viruses are used to generate realistic network traffic and intrusions. The dataset contains four types of attacks: botnet, C2, normal, and background. C2 is one of the APT stages. Argus is used to extract features from the extremely large PCAP files.

- **N-BaIoT**: Meidan et al. [104] establish the N-BaIoT dataset by configuring nine Internet-of-Things devices. Four types of attacks are used: reconnaissance, man-in-the-middle, denial-of-service, and botnet malware. Botnet attacks also include the establishment of a C2 channel; thus, reconnaissance, man-in-the-middle attacks, and the part of the botnet are considered APT attacks. 115 features are extracted from PCAP files including packets, steam, and time-frame features.

- **DS2OS**: DS2OS dataset is firstly introduced in [116] and generated using Distributed Smart Space Orchestration System (DS2OS) in a virtual IoT-integrated platform. DS2OS consists of several popular IoT-related intrusions. The whole architecture is a collection of different micro-services in an IoT environment that communicates via MQTT protocol. There are 347,935 normal observations and 10,017 malicious observations in this dataset. There are seven different types of attack traffic, including DoS, MC, MO, Probe, Scan, Spy, Wrong setup.

- **Edge-IIoTset** is a realistic dataset of IoT and IIoT studying which is applicable to design centralized MLs and federated learning algorithms. It consists of 14 threats which are grouped into 5 classes, including DoS/DDos, information gathering, man-in-the-middle, injection, and malware. Additionally, 61 additional features are extracted from a variety of sources (e.g., alarms, logs, network traffic) based on 1176 previously discovered features [49].

- **Aposemat IoT-23** is a dataset aiming for IoT applications, and it was introduced firstly in 2020, with data captured from 2018 to 2019. It is composed of data collected from twenty malware IoT devices and three normal IoT devices [139]. It is critical to note that the benign traffic was collected from real equipment (e.g., a LED lamp, an Amazon Echo home intelligent personal assistant, and a smart doorlock) rather than simulations.

- **MAWILab** dataset contains 4 classes including normal traffic, anomalous, suspicious, and notice. The dataset with numerous samples such as numeric values, discrete data, IP addresses, and identification numbers is widely used for anomaly detection research. Additionally, with the motivation of understanding anomaly detection, MAWILab makes use of two separate anomaly identification methods, including a straightforward predictive algorithm and a taxonomy of backbone traffic suspicious data [2].

Private IoT datasets are generated and applied for security research in several works [17, 58, 118, 130, 172]. To be specific, the authors in [130] indicate samples are extracted from a normal network scheme to generate a dataset for ML algorithms using. The private dataset consists of normal traffic, background and attack samples in [130]. Moreover, the study in [17] deploys dedicated sniffers to assemble network traffic as a dataset from network layer and MAC layer in a designed Open Systems Interconnection-based network. The authors in [172] firstly build an abnormal framework call dataset collected in four intrusion scenes. Meanwhile, the authors in [172] indicate people can contact them for accessing the private data using an email address left in the paper. Furthermore, the study in [118] utilizes a composed dataset with host features and flow features collecting data from sensors, network traffic, and system resource logs on a smart factory. In [58], the authors introduce a private dataset creation for benign and malware attacks separately. Benign samples are obtained from a system with Raspberry Pie II while malware samples are captured via monitoring ARM-based malware in a virus analysis system. However, most papers focus on performance of ML or deep learning algorithms in the topic about intrusion detection in IoT networks rather than introducing a private dataset creation.

Table 8. Summary of IoT Datasets. PA:Public Available

| Dataset | Year | PA | Attack Traffic | Meta-data | Format | Realistic |
|---------|------|----|----------------|-----------|--------|-----------|
| MAWILab [2] | 2010 | N | Attack, Special, Unknown | N | Tabular | Y |
| CTU-13 [53] | 2014 | Y | Botnet, C2, Background | Y | Packets, Malware | Y |
| N-BaIoT [104] | 2018 | Y | Bashlite, Mirai | N | Tabular | Y |
| DS2OS [116] | 2018 | Y | DoS, Malicious Control, Malicious Operation, Spying, Wrong Setup, Scan, Data Type Probing | N | Tabular | N |
| Bot-IoT [80] | 2019 | Y | DDoS, DoS, OS and Service Scan, Keylogging, Data exfiltration | Y | Packets, Tabular | Y |
| TON-IoT [8] | 2020 | Y | backdoor, ddos, dos, injection, mitm, password, ransomware, scanning, xss | Y | Packets, Tabular | Y |
| MQTTTest [123] | 2020 | Y | SlowITe, Bruteforce, Malformed data, Flooding, DoS attack | Y | Packets, Tabular | Y |
| Aposemat IoT-23 [139] | 2020 | Y | Botnet, C2, Background | Y | Packets, Flow | Y |
| Edge-IIoTset [49] | 2022 | Y | DoS/DDoS attacks, Information gathering, Man in the middle attacks, Injection attacks, Malware attacks | Y | Packets, Tabular | Y |

## 6 OPEN ISSUES, CHALLENGES AND OPPORTUNITIES AGAINST APT ATTACKS AND COMMON NETWORKS DETECTION

In this section, possible research directions are presented to increase the use of ML techniques to mitigate common issues and opportunities (e.g., updated attacks detection, IoT data characteristic, and ML algorithm selection and configuration [126, 143]) in attack detection and APT attacks open issues and opportunities (e.g., lack of dedicated APT attacks dataset, adversarial machine learning-based attacks and detection, and combing with malware detection) in IoT systems. Meanwhile, anomaly detection approach proves effectiveness in various attacks with encouraging performance while it is not applied widely for APT attacks. We demonstrate a promising opportunities for ATP attack detection using anomaly detection.

### 6.1 Common Challenges and opportunities in Network Intrusion

*6.1.1 Updated Attacks Detection.* Billions of electronic devices (e.g., sensors) and machines are being connected to IoT systems. This connectivity trend is anticipated to continue well into the future, especially regarding wireless IoT in applications such as smart cities. It means each piece of equipment added to the network has multiple chances to be a vector for Zero-day attacks. Therefore, ML algorithms should have the ability to learn adaptively and continuously to alleviate these anticipated attacks given the scale of network entry points.

The current ML paradigm works in two steps: first, a specific training dataset runs a machine-learning algorithm to construct a model, and then the model is used for its intended IoT application [36, 37]. The process of continuous ML is still in early development, so challenges exist in applying continuous learning to IoT. In [146], the authors demonstrate lifelong ML taking into account systems that can learn various tasks from one or more domains during their lifetime. The objective is to absorb learned sequentially to selectively apply that knowledge when learning a new scheme to form more refined assumptions and/or policies. Moreover, the authors suggest that the AI community should look beyond merely one-time learning algorithms to more seriously consider the nature of systems that can learn throughout their operational lifetime. Consequently, in the future, we can consider ways and means to deploy lifelong ML to identify cyber intrusion in the future in IoT.

*6.1.2    IoT Data Characteristic.* ML algorithms rely on data to train and generate an accurate model. Since the data are the basis of extracting knowledge, it is crucial to ensure the data's high fidelity. Data quality, availability, and integrity play a critical role in AI algorithms training and testing. However, IoT systems generate high volume, high velocity, and different data varieties, sometimes referred to as the 3 V's. Meanwhile, various types of equipment and devices are used to generate data that introduce data heterogeneity. Consequently, it is a challenge to ensure data authentication in IoT. However, ML algorithms show great promise in addressing these IoT security challenges.

*6.1.3    Machine Learning Algorithm Selection and Configuration.* AI algorithms each have particular characteristics, abilities, and limitations. Table 5, Table 6 and Table 7 describes individual model benefits and open issues. For instance, KNN-based approach performs high detection rate up to 99.99% using the DS2OS dataset while DAE indicates less features are applied in attack detection. There is currently no guideline on how to select AI algorithms for IoT. Therefore, users need to choose ML models carefully.

## 6.2    Challenges and Opportunities for APT Attack Detection

*6.2.1    Lack of Dedicated APT Network Intrusion Dataset.* There are a number of attractive datasets for network intrusion detection investigating. For instance, KDD 99 is one of the most utilized in studying network IDS [12, 47, 75]. The KDD 99 is public, that is a benchmark to evaluate performance between provided approached. However, there is no public dataset including pure ATP attacks. Some popular datasets (e.g., NSL-KDD ) contain several types of ATP attacks (e.g., U2R, R2L) while most of the malicious samples within the NSL-KDD dataset are DoS attacks. It is difficult to use them for APT studying, which is the biggest challenge for researching and analyzing. Furthermore, as APT attackers compromise hosts gradually, even packets sent by the same ports can have different labels. As APT attack records are already rare in the dataset, a detailed malware behavior analysis should be provided, so that the dataset can be labeled more carefully to reduce the false positive and enhance IDS's performance.

*6.2.2    Adversarial Machine Learning-based Attack and Detection.* One of the challenges of APT attacks is to bypass the defense systems without being discovered. Since Adversarial Machine Learning (AML) aims to mislead the ML classifiers, many of current works focus on employing AML on network intrusion to circumvent ML-based security systems [90]. Sriram et al. [150] demonstrate that adversarial samples created via AML approaches such as Jacobian-based Saliency Map Attack and Fast Gradient Sign Method can considerably affect machine/deep learning-based NIDS. While some research provide defense measures, for example, Ravi et al. [127] use a two-level DNN framework to detect fraudulent DNS domains generated by AML, there are still tough issues to address, such as increasing the speed and effectiveness of AML training.

*6.2.3    Combining with Malware Detection.* Due to the fact that APT attacks require the submission of malware in order to establish C2 and data exfiltration tunnels, numerous studies have focused on applying malware detection to the APT process; for example, Sriram et al. use an end-to-end deep learning approach to identify a variety of malware with varying file sizes [152]. As a result, merging the detection methods of both sectors could be a promising direction for the future.

## 7    SUMMARY

This survey article has reviewed current literature related to attacks and their countermeasures concerning the Internet of Things (IoT). More specifically, we have focused on ML-based methods currently being used to mitigate IoT network attacks. Following upon a general introduction to IoT and Industrial IoT (IIoT), security in IIoT, specifically Industrial Control Systems, has been described. Common attacks and APT attacks in an IoT network have been presented alongside An attack taxonomy and the APT attack distribution in this taxonomy. Based upon these attack types, PASTA threat model has been used to illustrate vulnerabilities and attack paths in IIoT networks. In response to various attack types, network intrusion detection methodologies have been summarized in the form of three categories: signature-based, anomaly-based, and hybrid approaches.

As a promising strategy to enhance the detection performance in mitigating attacks, ML-based methods have been presented by also focusing on anomaly-based and hybrid approaches to detect and classify attacks on IoT networks. ML algorithms have been studied and compared in terms of their detection performance under certain attack types. Finally, to help researchers who work in IoT, IIoT, ML and cybsersecurity, open issues and challenges for network intrusion have been discussed based on ML techniques alongside those for protecting IoT networks from APT attacks.

## ACKNOWLEDGMENTS

This work is supported in part by the Ontario Centre for Innovation under ENCQOR 5G Project #31993.